\documentclass[aps,floatfix,12pt]{iopart}
 \usepackage[pdftex]{color,graphicx}

 \newcommand{\be}{\begin{equation}}
 \newcommand{\ee}{\end{equation}}
 \newcommand{\bea}{\begin{eqnarray}}
 \newcommand{\eea}{\end{eqnarray}}

 \newcommand{\fnl}{f_{\rm NL}}
 \newcommand{\tnl}{\tau_{\rm NL}}
 \newcommand{\gnl}{g_{\rm NL}}
 
 \newcommand{\A}{{\cal A}}
 \newcommand{\B}{{\cal B}}
 \newcommand{\E}{{\cal E}}
 \newcommand{\C}{{\cal C}}
 \renewcommand{\H}{{\cal H}}
 \newcommand{\R}{{\cal R}}
 \newcommand{\bx}{{\bf x}}
 \newcommand{\bk}{{\bf k}}
 
 \newcommand{\bq}{{\bf q}}
 \newcommand{\bqp}{{\bf q'}}

 \newcommand{\bki}{{\bf {k_1}}}
 \newcommand{\bkii}{{\bf {k_2}}}
 \newcommand{\bkiii}{{\bf {k_3}}}
 \newcommand{\bkiv}{{\bf {k_4}}}

\begin{document}
\title{Local non-Gaussianity from inflation}

\author{David Wands}

\address{Institute of Cosmology \& Gravitation, Dennis Sciama Building, Burnaby Road,
University of Portsmouth, Portsmouth, PO1~3FX, United Kingdom\\
and\\
Yukawa Institute for Theoretical Physics, Kyoto University, Kyoto 606-8502, Japan}

\date{\today}

\begin{abstract}

The non-Gaussian distribution of primordial perturbations has the potential to reveal the physical processes at work in the very early Universe. Local models provide a well-defined class of non-Gaussian distributions that arise naturally from the non-linear evolution of density perturbations on super-Hubble scales starting from Gaussian field fluctuations during inflation. I describe the $\delta N$ formalism used to calculate the primordial density perturbation on large scales and then review several models for the origin of local primordial non-Gaussianity, including the cuvaton, modulated reheating and ekpyrotic scenarios.
I include an appendix with a table of sign conventions used in specific papers.

\end{abstract}
%


\section{Introduction}

The common presumption that primordial density perturbations have a Gaussian distribution is a powerful simplifying assumption that allows one to specify all the properties of the distribution once the two-point correlation function is known in real space, or equivalently the power spectrum in Fourier space. In particular the three-point and connected higher moments of the distribution vanish. On the other hand the statement that a distribution is non-Gaussian opens up an infinite array of possibilities. This has led to an assortment of empirical tests for non-Gaussianity of the primordial perturbations. By contrast there are relatively few non-Gaussian distributions that are motivated by theoretical models for the origin of structure in the early universe. While one can argue that a Gaussian distribution could describe density perturbations arising from a wide range of possible sources, any detection of deviations from a Gaussian distribution predicted by a specific theoretical model would be strong evidence in support of that model.

Vacuum fluctuations in light, weakly-coupled scalar fields during a period of inflation (defined here as accelerated expansion, $\ddot{a}>0$) in the very early universe provide a natural origin for an almost Gaussian distribution of field perturbations on large scales. Fluctuations in a free quantum field on small scales with comoving wavenumber, $k$, are swept up to scales much larger than the comoving Hubble scale, $H^{-1}/a=1/\dot{a}$, which shrinks during inflation. On super-Hubble scales ($k<aH$) damping drives the fluctuations into a squeezed state and they can effectively be treated as a classical random distribution \cite{Polarski:1995jg}. Models of slow-roll inflation in the early universe are driven by canonical fields whose self-interactions are small relative to the Hubble scale and hence their distribution on scales close to the Hubble scale remains well described by a Gaussian distribution with vanishing third- and higher-order connected functions \cite{Maldacena:2002vr,Seery:2005gb}.

As inflation continues to stretch the perturbations up to scales far larger than the Hubble scale, spatial gradients are expected to become negligible and the we may treat the evolution locally as effectively ``separate universes'' \cite{Salopek:1990jq,Wands:2000dp} whose initial conditions are set by the Gaussian distribution of the scalar field values during inflation. This motivates the study of local models of primordial non-Gaussianity where the distribution of primordial density perturbations, $\zeta$, can be described by a local function of one or more Gaussian random fields, $\zeta(\delta\varphi^I)$. A linear function of Gaussian fields is itself Gaussian, so a non-Gaussian distribution implies non-linearity and the study of non-linear cosmological perturbations. This local model for non-Gaussianity turns out to be a very good description of non-Gaussianity in some simple physical models for the origin of structure in the very early universe.

Conventionally the primordial perturbations are characterised by the 
metric potential in the matter-dominated era, $\Phi=(3/5)\zeta$.
In Fourier space we define the power spectrum and bispectrum as
\bea
\langle\Phi_{\bki}\Phi_{\bkii} \rangle = (2\pi)^3 P_{\Phi}(k_1) \delta^{3}(\bki+\bkii) \,,\\
\label{bispectrum}
\langle \Phi_{\bki} \Phi_{\bkii} \Phi_{\bkiii} \rangle = (2\pi)^3 B_{\Phi}(k_1,k_2,k_3) \delta^{3}(\bki+\bkii+\bkiii) \,,
\eea
and the amplitude of the bispectrum relative to the power spectrum is then given by the dimensionless parameter
\be
 \label{deffnl}
\fnl(k_1,k_2,k_3)
 \equiv \frac{B_{\Phi}(k_1,k_2,k_3)}{2 \left[ P_{\Phi}(k_1) P_{\Phi}(k_2) + P_{\Phi}(k_2) P_{\Phi}(k_3) + P_{\Phi}(k_3) P_{\Phi}(k_1) \right]} \,.
\ee

In a perturbative expansion, the second-order expression for the 
metric potential as a local function of a single Gaussian field, $\phi$, is given by~\cite{Komatsu:2001rj}
 \be
  \label{originalPhi}
 \Phi(\bx) = \phi(\bx) + \fnl \left( \phi^2(\bx) - \langle \phi^2 \rangle \right) +\ldots \,.
 \ee
In this special case $\fnl$ is, by construction, a constant parameter independent of spatial position or scale. However in more general local models the ratio of the bispectrum to the power spectrum given by Eq.~(\ref{deffnl}) may be taken as a definition of $\fnl$ which is then scale and shape-dependent.
Note that following Komatsu and Spergel~\cite{Komatsu:2001rj} we adopt a sign convention for the 
metric potential $\Phi$, and hence $\fnl$, which is the opposite of that used by, for example, Mukhanov et al~\cite{Mukhanov:1990me} and Maldacena \cite{Maldacena:2002vr}. 
Specific sign conventions are summarised in a table in an Appendix.

In section 2 I review the $\delta N$-formalism that is commonly used to calculate the primordial density perturbation on large scales and its higher-order correlations, including the use of Feynman-type diagrams and how the $\delta N$ formalism extends to the description of primordial isocurvature density perturbations. In section 3 I review several examples of models for the origin of structure in the very early universe and the local non-Gaussianity they give rise to. I conclude in section 4 with brief review of current observational constraints.

\section{The $\delta N$ expansion}

A powerful technique for calculating the non-linear primordial density perturbation in many cosmological models is the $\delta N$ formalism \cite{Starobinsky:1986fxa,Sasaki:1995aw,Wands:2000dp,Lyth:2005fi}, which identifies the primordial density perturbation with the perturbed logarthmic expansion, $N=\int Hdt$.

In Friedmann-Robertson-Walker (FRW) cosmology there is a preferred foliation of spatial hypersurfaces which are maximally symmetric, and on which the matter density and pressure are also homogeneous and isotropic. In an inhomogeneous universe we can use the difference between uniform-expansion hypersurfaces and uniform-matter hypersurfaces as a measure of inhomogeneity.
%
At first order we define \cite{Bardeen:1988,Malik:2008im}
 \be
 \zeta_1 \equiv
 \C_1 - \frac{H}{\dot\rho}\delta_1\rho \,.
 \ee
where $\C_1$ and $\delta_1\rho$ are the gauge-dependent spatial metric and density perturbations respectively \cite{Mukhanov:1990me,Malik:2008im}
(see Appendix~A).
This can be interpreted either as a curvature perturbation on uniform-density hypersurfaces (where $\delta\rho=0$) or a dimensionless density perturbation on uniform-curvature hypersurfaces (where $\C=0$).
%

Local energy conservation equation \cite{Wands:2000dp} ensures that $\zeta_1$ remains constant for adiabatic density perturbations in the long-wavelength limit where $(k/aH)^2\Phi_1\to0$ and $\zeta_1$ approaches the comoving curvature perturbation,
$\R$.
Thus during inflation it is convenient to evaluate the comoving curvature perturbation on large scales, $k/aH\to0$, in order to determine the primordial density perturbation from single-field inflation
 \be
 \zeta_1 \to \R_1 = \C_1 - \frac{H}{\dot\varphi}\delta_1\varphi \,.
 \ee
Because the non-adiabatic decaying mode rapidly decays outside the Hubble scale \cite{Gordon:2000hv} it is sufficient to evaluate the primordial perturbation due to scalar field perturbations shortly after Hubble exit, $k=aH$.
However in multi-field inflation, additional light fields can lead to non-adiabatic pressure perturbations on super-Hubble scales and $\zeta_1$ becomes time-dependent in general
\cite{Starobinsky:1994mh,GarciaBellido:1995qq}.

Beyond linear order,
we can identify the non-linear metric perturbation which coincides with $\zeta_1$ at first order and on large scales \cite{Lyth:2004gb} (see also \cite{Rigopoulos:2003ak,Langlois:2005ii} and the article by Langlois and Vernizzi elsewhere in this issue~\cite{Langlois:2010CQG})
 \be
  \label{defzeta}
 \zeta(t,\bx) = \delta N(t,\bx) + \frac13 \int_{\bar\rho(t)}^{\rho(t,\bx)} \frac{d\tilde\rho}{\tilde\rho+\tilde{P}} \,.
 \ee
where $\bar\rho(t)$ is the homogenous background density, $\rho(t,\bx)$ denotes the local inhomogeneous density and the local expansion is given by $N=\int {H}dt$, where ${H}(t,\bx)=\nabla_\mu u^\mu/3$ is the local Hubble expansion rate along comoving worldlines, $u^\mu=dx^\mu/d\tau$.

A local form for the distribution of $\zeta$ naturally arises from the evolution of matter fields on scales much larger than the Hubble scale in the very early universe. After averaging on some scale, $L\gg H^{-1}$, such that spatial gradients and anisotropy can be neglected, then the local expansion is well-described by the Friedmann equation for a homogeneous universe with the corresponding local density and pressure. This is known as the ``separate universe'' approach \cite{Wands:2000dp}, and can be derived from the full inhomogeneous equations of motion as a long-wavelength limit in a gradient expansion \cite{Salopek:1990jq,Rigopoulos:2003ak,Langlois:2005ii}. In this approach the local evolution, along a given worldline, is determined by the initial values of the matter fields on an initial spatial hypersurface.

Therefore one can evaluate $\zeta(t_f,\bx)$, defined by Eq.~(\ref{defzeta}), as $\delta N$, the perturbed expansion on a uniform-density hypersurface ($\rho(t_f,\bx)=\bar\rho(t_f)$ for all $\bx$) at some final time $t_f$ after inflation has ended by evaluating the integrated expansion from some initial spatially flat hypersurface, $\delta N(t_i,\bx)=0$ for all $\bx$ \cite{Starobinsky:1986fxa,Sasaki:1995aw,Lyth:2005fi}. In the separate universe framework \cite{Wands:2000dp} the inhomogeneous universe is modelled as a patchwork of locally homogeneous regions which, due to causality, evolve independently. In particular one can determine the non-linear local expansion $N$ using the Friedmann equation for the local expansion as a function of the initial local fields, $\varphi^I(t_i,\bx)=\bar\varphi(t_i)+\delta\varphi(t_i,\bx)$. Assuming $N(\varphi^I)$ is an analytic function of the initial field values, we have the Taylor series expansion \cite{Lyth:2005fi}
 \be
 \label{zetaTaylor}
 \zeta = N(\varphi^I) - \bar{N} = N_I \delta\varphi^I + \frac12 N_{IJ} \delta\varphi^I \delta\varphi^J + \frac16 N_{IJK} \delta\varphi^I \delta\varphi^J \delta\varphi^K + \ldots \,,
 \ee
where $N_I=\partial N/\partial\varphi^I$, etc, and summation is implied over repeated indices. If the field perturbations themselves are constructed as an expansion in a small perturbation parameter
 \be
 \varphi^I = \bar\varphi^I +\delta_1\varphi^I +\frac12 \delta_2\varphi^I + \frac16 \delta_3\varphi^I +\ldots \, ,
 \ee
then we can write, order by order in a perturbative expansion $\zeta=\zeta_1+\zeta_2/2+\zeta_3/6\ldots$, where
 \bea
 \label{zeta1}
 \zeta_1 = N_I \delta_1\varphi^I \,, \\
 \label{zeta2}
 \zeta_2 = N_{IJ} \delta_1\varphi^I \delta_1\varphi^J + N_I \delta_2\varphi^I \,,\\
 \label{zeta3}
 \zeta_3 = N_{IJK} \delta_1\varphi^I \delta_1\varphi^J \delta_1\varphi^K + 3 N_{IJ} \delta_1\varphi^I \delta_2\varphi^J + N_I \delta_3\varphi^I \,,
 \eea

To calculate the statistics of the primordial perturbation, $\zeta$, we therefore need to know the dependence of the large scale expansion upon the initial field values, $N(\varphi^I)$, and the statistical distribution of the initial field values. The distribution of the fields is determined by the quantum fluctuations of the vacuum in any given inflationary model. We take the first-order perturbations to describe the vacuum fluctuations of the non-interacting free fields, and build up the higher-order terms from the interaction Hamiltonian, following Maldacena \cite{Maldacena:2002vr} (see also the article by Koyama elsewhere in this issue \cite{Koyama:2010xj}).

In canonical slow-roll inflation the linear field perturbations defined on unperturbed spatially flat hypersurfaces \cite{Mukhanov:1988jd,Sasaki:1986hm} are given by the flat spacetime vacuum on sub-Hubble scales and are then approximately constant on super-Hubble scales with power spectra
 \be
 \langle \delta_1\varphi^I_\bki \delta_1\varphi^J_\bkii \rangle \simeq (2\pi)^3 P(k_1) \delta^{IJ} \delta^3(\bki+\bkii) \,,
 \ee
where $P(k_1)\approx H_*^2/2k_1^3$ and $H_*$ is the Hubble scale at Hubble-exit, $k=a_*H_*$. Note that the dimensionless power spectrum, related to the variance in real space, is multiplied by a volume factor in $k$-space to give
 \be
 {\cal P}(k) = \frac{4\pi k^3}{(2\pi)^3} P(k) \simeq \left( \frac{H_*}{2\pi} \right)^2 \,,
 \ee
and this is approximately scale invariant, with spectral tilt
\be
 \left| \frac{d\ln {\cal P}}{d\ln k} \right| \ll 1 \,.
 \ee

Scalar field interactions are suppressed by slow-roll parameters and so field perturbations on scales close to the Hubble scale are approximately Gaussian. In the following we will assume that all connected higher-order moments of the scalar field perturbations vanish at some fixed time during inflation, shortly after all relevant scales have crossed outside the Hubble scale, $k<aH$. Hence $N$ is well-described by a local function of Gaussian random fields soon after the smallest relevant modes have left the Hubble scale \footnote{Alternatives to slow-roll models of inflation based on non-canonical scalar field Lagrangians with non-linear kinetic terms, such as k-inflation \cite{ArmendarizPicon:1999rj} or DBI inflation \cite{Alishahiha:2004eh}, can lead to significant non-Gaussianity of the field perturbations at Hubble exit and hence produce different types of non-Gaussianity, different from the local form.}.

We will assume that $\delta_2\varphi^I$, $\delta_3\varphi^I$, etc, vanish sufficiently close to Hubble exit and thus the higher-order moments are given by the first terms on the right-hand-side of Eqs.~(\ref{zeta2}) and~(\ref{zeta3}). In particular the primordial power spectrum is given, at leading order, by
 \be
 \label{treePzeta}
 \langle \zeta_\bki \zeta_\bkii \rangle = N_I N_J \langle \delta\varphi^I_\bki \delta\varphi^J_\bkii \rangle = (2\pi)^3 P_\zeta(k_1) \delta^3(\bki+\bkii) \,,
 \ee
where
 \be
 P_\zeta(k_1) = N_I N^J \delta^{IJ} P(k_1) \,,
 \ee
In the next subsection we will discuss higher-order contributions to the primordial power spectrum coming from higher-order terms in the $\delta N$ expansion. These can be identified as ``loop'' corrections in a diagrammatic approach \cite{Byrnes:2007tm}, where the leading order result (\ref{treePzeta}) is the ``tree-level'' term.

Similarly we can construct the leading order terms in the bispectrum and trispectrum
 \bea
 \langle \zeta_\bki \zeta_\bkii \zeta_\bkiii \rangle = (2\pi)^3 B_\zeta(k_1,k_2,k_3) \delta^3(\bki+\bkii+\bkiii) \,,\\
 \langle \zeta_\bki \zeta_\bkii \zeta_\bkiii \zeta_\bkiv \rangle = (2\pi)^3 T_\zeta(k_1,k_2,k_3) \delta^3(\bki+\bkii+\bkiii+\bkiv) \,,
 \eea
where
 \bea
 \label{Bzeta}
 B_\zeta(k_1,k_2,k_3) &=& N_I N_J N^{IJ} \left[ P(k_1) P(k_2)
 + 2~{\rm perms}
 \right] \,,\\
 \label{Tzeta}
 T_\zeta(k_1,k_2,k_3) &=& N_{IJK} N^I N^J N^K \left[ P(k_2) P(k_3) P(k_4)
 + 3~{\rm perms}
 \right] \nonumber\\
&& +   N_{IJ}N_K^J N^I N^K \left[ P(k_3) P(k_4) P(k_{13}) + 11~{\rm perms} \right] \,,
 \eea
and $k_{12}=|\bki-\bkii|$, etc.

\subsection{Diagrammatic approach}

The perturbative expression of the non-linear expansion as a function of the field values during inflation (\ref{zetaTaylor}) is conveniently expressed in terms of Feynman-type diagrams \cite{Byrnes:2007tm} (see also \cite{Zaballa:2006pv}). Similar diagrams have been used in the analysis of large-scale structure in cosmology \cite{Scoccimarro:1995if,Crocce:2005xy}. Every term in the $n$-point function can be identified with a diagram with $n$ external lines. To perturbative order $r$ one should draw all diagrams with $r$ internal propagators. Tree-level diagrams correspond to $r=n-1$ while diagrams with $r\geq n$ are loop corrections. If the expansion is a local function of Gaussian fields then the internal propagators have no self-interactions. Examples are shown in Figure~1. A comprehensive prescription for drawing diagrams and constructing the corresponding terms to any given order are given by Byrnes et al \cite{Byrnes:2007tm}.

\begin{figure}
\centering
\includegraphics[width=1\textwidth]{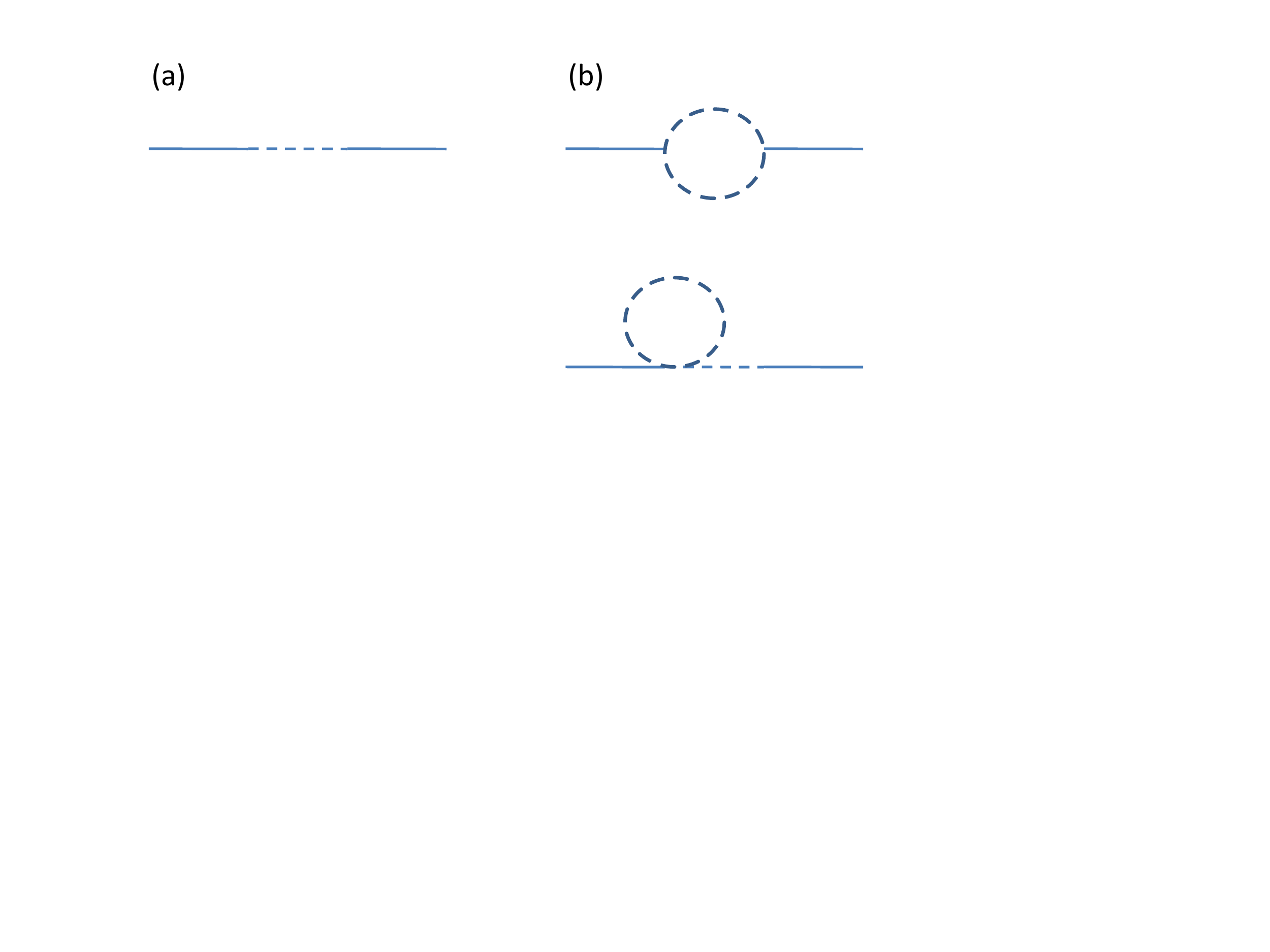}
\label{diagrams}
\vspace*{-3in}
\caption{Power spectrum diagrams for $\delta N$ expansion with Gaussian fields (a) at tree level, and (b) one loop, corresponding to each term in Eq.~(\ref{P1loop}).}
\end{figure}

Here we will consider only diagrams for Gaussian fields, $\varphi^A$. The power spectrum up to one-loop is then shown in figure~1, corresponding to
\begin{eqnarray}
 \label{P1loop}
P_\zeta(k) = N_A N_B \delta^{AB} P(k)\nonumber\\
\quad \ + \frac{1}{(2\pi)^3} \int d^3\bq \, d^3\bqp \left( \frac12 N_{AB}N_{CD} \delta^3(\bqp-\bq+\bk) +N_{A}N_{BCD} \delta^3(\bqp-\bk) \right)
 \nonumber \\
  \qquad \qquad \qquad \delta^{AC} \delta^{BD} P(q) P(q') \,.
\end{eqnarray}
Similarly the bispectrum and higher-order correlations can be evaluated including an arbitrary number of loop corrections.


Because the loop corrections include integrations over all internal momenta, and the scalar field perturbations are expected to have almost scale invariant power spectra during inflation, the loop corrections can give corrections that, although only slowly growing on large scales, formally diverge in the infra-red limit. This issue is discussed by Seery elsewhere in this issue~\cite{Seery:CQG}. Typically the corrections remain small in a region which is not much larger than the observable universe, but in some cases, for instance if the tree-level terms are absent~\cite{Boubekeur:2005fj}, then loop corrections may play an important role.

Some loop corrections, such as the second one-loop term in Eq.~(\ref{P1loop}), can be identified as dressing the vertices. Byrnes et al \cite{Byrnes:2007tm} noted that these can be interpreted as renormalising the derivatives of $N$ in terms of the average local derivatives:
\begin{equation}
\langle\tilde{N}_A\rangle  \equiv \int d^3q N_A + \frac{1}{(2\pi)^3} \int d^3q N_{ABC} \delta^{BC} P(q) + \ldots
\end{equation}
Renormalising the vertices in this way simplifies the expressions for the loop corrections, to yield only one new term at each new order. In real space this is sufficient to resum all divergences, but in Fourier space the dependence on IR wavenumbers on the convolution remains.
For instance the power spectrum
to one loop becomes
\begin{eqnarray}
P_\zeta(k) = \langle\tilde{N}_A\rangle \langle \tilde{N}_B\rangle \delta^{AB} P(k) \nonumber\\
\quad + \frac{1}{(2\pi)^3} \int d^3\bq \frac12 \langle\tilde{N}_{AB}\rangle\langle\tilde{N}_{CD}\rangle \delta^{AC} \delta^{BD} P(k) P(k-q) \,.
\end{eqnarray}

\subsection{Single-field vs multi-variate local models}

If the primordial density perturbation (\ref{zetaTaylor}) is a local function of a single scalar field
\be
 \zeta = N'\delta\varphi + \frac12 N''\delta\varphi^2 + \frac16 N'''\delta\varphi^3+\ldots \,,
\ee
then the expressions for the higher order moments of the primordial distribution, e.g., (\ref{Bzeta}) and (\ref{Tzeta}), simplify considerably and we can write
 \bea
 B_\zeta(k_1,k_2,k_3) &=& \frac65 \fnl \left[ P_\zeta(k_1) P_\zeta(k_2)
 + 2~{\rm perms}
 \right] \,,\\
 T_\zeta(k_1,k_2,k_3) &=& \frac{54}{25} \gnl \left[ P_\zeta(k_2) P_\zeta(k_3) P_\zeta(k_4)
 + 3~{\rm perms}
 \right] \nonumber\\
&& +   \tnl \left[ P_\zeta(k_3) P_\zeta(k_4) P_\zeta(k_{13}) + 11~{\rm perms} \right] \,,
 \eea
where at leading order (tree-level) the non-linearity parameters are given by~\cite{Lyth:2005fi,Byrnes:2006vq}
\be
\fnl = \frac56 \frac{N''}{N^{\prime2}}
\,, \qquad \tnl = \frac{36}{25} \fnl^2
\,, \qquad \gnl = \frac{25}{54} \frac{N'''}{N^{\prime3}} \,.
\ee

The $\delta N$ expression for the primordial perturbation (\ref{zetaTaylor}) goes beyond the simplest local model (\ref{originalPhi}) by considering the case where the primordial perturbation is a function of more than one Gaussian random field. Including terms up to second order we have
\be
 \Phi(x) = \sum_I \Phi^I(x) + \sum_{I,J} f_{IJ} \left( \Phi^I(x)\Phi^J(x) - \langle\Phi^I\Phi^J\rangle \right) +\ldots\,,
 \ee
where we have defined
 $\Phi^I=(3/5)N_I\delta\varphi^I$ and $f_{IJ}=(6/5)N_{IJ}/N_IN_J$ (no sums).
The non-linearity parameter $\fnl$ defined by Eq.~(\ref{deffnl}) is then $k$-dependent and given by~\cite{Byrnes:2009pe}
\be
 \fnl = \frac{\sum_{I,J} N_{IJ} \left[ P_I(k_1)P_J(k_2) + P_I(k_2)P_J(k_3) + P_I(k_3)P_J(k_1) \right]}{\sum_{I,J} N_IN_J \left[ P_I(k_1)P_J(k_2) + P_I(k_2)P_J(k_3) + P_I(k_3)P_J(k_1) \right]} \,.
\ee
This is a constant if the scale-dependence of all the fields which contribute to the primordial perturbation is the same, but in general it is scale and shape dependent. $k$-dependence of $\fnl$ defined by Eq.~(\ref{deffnl}) also arises due to interactions and hence non-Gaussianity of fields on super Hubble scales~\cite{Byrnes:2009pe}, but this goes beyond the strict definition of local models considered here.
If the fields are all approximately massless during inflation then their scale-dependence and interactions are small and the $k$-dependence of $\fnl$ is expected to be weak (but see \cite{Byrnes:2008zy}).

\subsection{Isocurvature non-Gaussianity}

Having identified the non-linear primordial density perturbation, $\zeta$, with the perturbed expansion, $\delta N$, up to a uniform density hypersurface, it is straightforward to extend this to $n$ primordial density perturbations in an $n$-component system, where we define
\begin{equation}
 \label{defzetaalpha}
 \zeta_\alpha \equiv \delta N + \frac13 \int_{\bar\rho_\alpha}^{\rho_\alpha} \frac{d\rho_\alpha}{\rho_\alpha+P_\alpha} \,.
\end{equation}

Relative density perturbations between the different components which leave the total energy density unperturbed are know as isocurvature perturbations. In particular a relative perturbation between photons and baryons around the time of last scattering of the CMB photons corresponds to a perturbation of the local entropy (determined by the photon number density) per baryon and hence such perturbations are also known as entropy perturbations. At linear order we have
\begin{equation}
S_B = \frac{\delta n_B}{n_B} - \frac{\delta n_\gamma}{n_\gamma} \,.
\end{equation}
We can identify this with the difference between the two primordial density perturbations $\zeta_B$ and $\zeta_\gamma$, up to a factor of 3. Hence beyond linear order we define \cite{Kawasaki:2008sn,Langlois:2008vk}
\begin{equation}
 \label{defSB}
S_B \equiv 3 \left( \zeta_\gamma - \zeta_B \right) \,.
\end{equation}
A similar definition is used for the cold dark matter isocurvature perturbation, $S_{CDM}$, and the neutrino isocurvature perturbation, $S_\nu$, relative to the photon number density.

More generally for an $n$ component system we can decompose an arbitrary density perturbation into one adiabatic density perturbation, $\zeta=\zeta_\alpha$ for all components $\alpha$, and $n-1$ independent isocurvature modes
\begin{equation}
S_{\alpha\beta} \equiv 3(\zeta_\alpha-\zeta_\beta) \,.
\end{equation}


One can define non-linearity parameters for the isocurvature bispectrum at leading order analogous to the adiabatic bispectrum (\ref{deffnl})
\be
 \label{deffiso}
f_S(k_1,k_2,k_3)
 = \frac{B_{S}(k_1,k_2,k_3)}{2 \left[ P_{S}(k_1) P_{S}(k_2) + P_{S}(k_2) P_{S}(k_3) + P_{S}(k_3) P_{S}(k_1) \right]} \,.
\ee
If the isocurvature modes are uncorrelated with the adiabatic density perturbation, and we use the linear relation between primordial matter isocurvature perturbations and the Newtonian potential, $\Phi=S_{CDM}/5$, we obtain \cite{Kawasaki:2008sn}
\be
 \fnl = \frac{5f_S}{162} \frac{P_{S}(k_1) P_{S}(k_2) + P_{S}(k_2) P_{S}(k_3) + P_{S}(k_3) P_{S}(k_1)}{P_{\Phi}(k_1) P_{\Phi}(k_2) + P_{\Phi}(k_2) P_{\Phi}(k_3) + P_{\Phi}(k_3) P_{\Phi}(k_1)} \,.
 \ee
However one should be wary of interpreting bounds on the adiabatic $\fnl$ as any constraint on isocurvature non-Gaussianity which leads to a distinctive bispectrum on the CMB
sky~\cite{Kawasaki:2008sn,Hikage:2008sk}. Moreover the isocurvature modes are in general correlated with adiabatic modes \cite{Langlois:1999dw} and hence there may exist non-vanishing cross-correlations, $\langle S\zeta\zeta\rangle$ and $\langle SS\zeta\rangle$ \cite{Langlois:2008vk}.

\section{Models of non-Gaussianity}

\subsection{Single inflaton}

A homogeneous scalar field cosmology evolves like that of a fluid with time dependent density and pressure, but an inhomogeneous scalar field is not in general a barotropic fluid \cite{Arroja:2010wy}. However the non-adiabatic part of the field's pressure perturbation is proportional to the field's comoving density perturbation \cite{Gordon:2000hv}. The Einstein equations energy and momentum constraints require that the comoving density perturbation vanishes in the large scale limit where the Newtonian potential stays finite \cite{Malik:2008im} and hence if the local energy density and pressure during inflation is dominated by a single scalar field then perturbations must be adiabatic in this large scale limit. The field perturbations enter a squeezed state in phase space and the local time derivative of the field is no longer independent of the local field value.

This makes it straightforward to predict the primordial density perturbation and hence the local non-Gaussianity for inflation models with a single inflaton field. As all perturbations are adiabatic then the local density and pressure follow the same phase-space trajectory in phase-space as the background cosmology with $\delta p/\delta\rho=\dot{p}/\dot\rho$. This ensures that the perturbation $\zeta$ is non-linearly conserved in this large scale limit \cite{Wands:2000dp,Lyth:2003im,Rigopoulos:2003ak,Lyth:2004gb,Langlois:2005ii} and we can calculate the primordial density perturbation, long after inflation, in terms of quantities during inflation. Using the $\delta N$-formalism we just need to calculate the effect of field perturbations on the local expansion, $N$, during inflation.

To first order we have
\be
 \zeta_1 = N' \delta\varphi = -\frac{H}{\dot\varphi} \delta\varphi \,,
\ee
where $\delta\varphi$ denotes field perturbations in the spatially flat gauge.
This is time-independent on large scales, but in practice it is convenient to evaluate it soon after Hubble exit ($k=aH$) where the amplitude of the field perturbations are given by $P(k)\simeq H_*^2/2k^3$. We then obtain the simple result for the primordial power spectrum at leading order
\be
 P_\zeta(k) = \left( \frac{H}{\dot\varphi} \right)_*^2 P(k) \,.
\ee
At second order, assuming a Gaussian field $\delta\varphi$, we have
\bea
 \zeta_2 = N'' \delta\varphi^2 = \left( -\frac{\dot{H}}{\dot\varphi^2} + \frac{H\ddot\varphi}{\dot\varphi^3} \right) \delta\varphi^2 \,,
 \eea
In terms of the usual slow-roll parameters $\epsilon=-\dot{H}/H^2$ and $\eta=-(\ddot\varphi/H\dot\varphi)-(\dot{H}/H^2)$ we have
\be
 \zeta_2 = \left( 2\epsilon - \eta \right) \zeta_1^2 \,.
 \ee
Thus we see that the second order primordial density perturbation is suppressed with respect to the square of the first-order perturbations and hence $|\fnl|\ll 1$ \cite{Falk:1992sf,Gangui:1993tt}. In fact at this order in the slow-roll approximaton we can no longer neglect the intrinsic non-Gaussianity of the fields at Hubble exit \cite{Maldacena:2002vr}, and a simple argument where we consider long-wavelength perturbations to set the local Hubble expansion as shorter wavelengths leave the Hubble scale \cite{Allen:2005ye} shows that in the squeezed limit, where $k_1^2\ll k_2^2+k_3^2$, we find
\begin{equation}
\fnl(k_1,k_2,k_3) = \frac56 \left( 3\epsilon-\eta \right) \,.
\end{equation}
More generally, for any single field model producing adiabatic perturbations during inflation the bispectrum can be related to the scale-dependence of the power spectrum in the squeezed limit and we have \cite{Creminelli:2004yq}
\begin{equation}
 \label{adiabaticlimit}
\fnl(k_1,k_2,k_3) = \frac5{12} (1-n_\zeta) \,.
\end{equation}
where $n_\zeta-1=d\ln {\cal P}_\zeta/d\ln k$. Observational bounds on the spectral tilt, $|n_\zeta-1|\ll1$, thus ensure that local-type non-Gaussianity is very small in single-inflaton field models.

\subsection{Curvaton}

The curvaton scenario~\cite{Mollerach:1989hu,Linde:1996gt,Enqvist:2001zp,Lyth:2001nq,Moroi:2001ct} provides a simple model for the origin of the primordial perturbations which could
exhibit significant non-Gaussianity of the local type~\cite{Lyth:2002my}.

The curvaton is a weakly-coupled scalar field which is light during inflation, $m\ll H$, but it has a negligible energy density at that time. Quantum fluctuations of the field during inflation generate an almost scale invariant spectrum of field perturbations on super-Hubble scales. We assume that the weakly-coupled field remains decoupled from the inflaton and its decay products at the end of inflation. The field begins to oscillate about the minimum of its potential when the Hubble rate drops below the mass of the field some time after inflation. An oscillating massive field, $m\gg H$, has a pressureless equation of state (averaged over several oscillation times) and hence the curvaton energy density decays as $a^{-3}$, but grows relative to the energy density of radiation, $\propto a^{-4}$. This is the Polonyi or moduli problem of weakly coupled scalar fields which can come to dominate the energy density of the early universe, disrupting the conventional hot big bang. This is not a problem so long as the moduli decay and their decay products thermalise before the epoch of primordial nucleosynthesis. The inhomogeneous density of the curvaton is transfered to the radiation when the curvaton decays leading to a primordial density perturbation on super-Hubble scales.

Note that, in contrast to inflaton perturbations, the curvaton field fluctuations are isocurvature field perturbations during inflation and thus give rise to non-adiabatic pressure perturbations after inflation which leads to a change in the value of the perturbation, $\zeta$, on super-Hubble scales. Thus the bispectrum in the squeezed limit is not constrained by the relation (\ref{adiabaticlimit}) for perturbations generated by the adiabatic mode during inflation.

In the simplest example of a curvaton field, $\chi$, with potential $V(\chi)=m^2\chi^2/2$, the curvaton field is a free field during inflation with no self-interactions and hence is described by a Gaussian field. The quadratic potential naturally leads to a primordial density field that is a quadratic local function of the Gaussian curvaton field and hence the simplest curvaton model is well described by the simplest local model of non-Gaussianity given in Eq.~(\ref{originalPhi}).

The
density perturbation for the oscillating curvaton field
on spatially flat hypersurfaces is given by
 Eq.~(\ref{defzetaalpha})
\begin{equation}
 \zeta_\chi = \frac13 \frac{\rho_\chi-\bar\rho_\chi}{\bar\rho_\chi} = \frac{2\bar\chi\delta\chi+\delta\chi^2-\langle\delta\chi^2\rangle}{\bar\chi^2} \,.
\end{equation}
In the case where the curvaton density remains sub-dominant throughout we can assume a linear transfer to the primordial density perturbation, with transfer efficiency $r\sim \Omega_{\chi,{\rm decay}}=\rho_\chi/\rho_{\rm tot}|_{\rm decay}$, and we can write
\begin{equation}
\zeta = r \zeta_\chi = \zeta_1 + \frac56 \fnl \left( \zeta_1^2 - \langle\zeta_1^2\rangle \right) \,.
\end{equation}
where we identify
\begin{equation}
 \label{defr}
\zeta_1 = \frac{2r}{3} \frac{\delta\chi}{\bar\chi} \,.
\end{equation}
and the non-linearity parameter
\cite{Lyth:2002my}
\begin{equation}
 \label{largefnlr}
\fnl = \frac{5}{4r} \,.
\end{equation}
For $r\ll 1$ we can have large, positive $\fnl$\footnote{There has been some confusion in the literature over the ``correct'' sign for $\fnl$ in the curvaton scenario.
%
See Appendix~A for a table of sign conventions used in specific papers.
In my own papers, the original result for $\fnl$ in Ref.~\cite{Lyth:2002my} had the ``correct'' sign (that used here, which coincides with that used for observational constraints \cite{Komatsu:2001rj}) due to a combination of the sign convention used and a sign error in the arithmetic. The mistake in the arithmetic was corrected, for instance in Ref.~\cite{Bassett:2005xm}, but then the sign was ``wrong'' because of the sign convention used.}. In this case current observational bounds on $\fnl$ \cite{Komatsu:2010fb} place a lower bound on $r>0.01$.

For $r\sim 1$ we need to include the gravitational effect of the curvaton density when it decays. If we assume an instantaneous decay hypersurface given by the local Hubble rate $H=\Gamma$, we obtain the non-linear relation between the curvaton perturbation $\zeta_\chi$ and the primordial perturbation $\zeta$~\cite{Sasaki:2006kq}
\be
 (1-\Omega_{\chi,{\rm decay}}) e^{-4\zeta} + \Omega_{\chi,{\rm decay}} e^{3(\zeta_\chi-\zeta)} = 1 \,,
\ee
from which we recover Eq.~(\ref{defr}) at linear order, where
$r=3\Omega_{\chi,{\rm decay}}/(4-\Omega_{\chi,{\rm decay}})$, and at second and third order we obtain the non-linearity parameters as a function of $r$
\cite{Bartolo:2003jx,Lyth:2005fi,Sasaki:2006kq}
\bea
\fnl = \frac{5}{4r} \left( 1-\frac{4r}{3} -\frac{2r^2}{3} \right) \,,
 \\
\gnl = -\frac{25}{6r} \left( 1 -\frac{r}{18} -\frac{10r^2}{9} - \frac{r^3}{3} \right)
 \,.
\eea
A measurement of both $\fnl$ and $\gnl$ would be an important consistency test of the simplest curvaton model as both are a function of a single parameter, $r$.
For a sub-dominant curvaton, $r\ll1$, we recover Eq.~(\ref{largefnlr}) for $\fnl$ and $\gnl\simeq-10\fnl/3$~\cite{Sasaki:2006kq}. In the opposite limit, as $r\to1$, where the curvaton dominates the energy density of the universe before it decays, we have $\fnl\to-5/4$ and $\gnl\to25/12$.

Other possibilities that have been studied include a kinetic-energy dominated era (rather than radiation dominated) before curvaton decay \cite{Giovannini:2003jw} and multiple curvaton-type fields \cite{Choi:2007fya}. Even in the case of multiple curvaton fields contributing to the primordial density perturbation we recover the same lower bound $\fnl\geq-5/4$ as in the single field case \cite{Assadullahi:2007uw,Huang:2008rj}.
A much wider range of local non-Gaussianity becomes possible if we consider the effect of self-interaction terms in the curvaton potential, such that $V=m^2\chi^2/2+\lambda\chi^n+\ldots$ with $n\geq3$ which can lead to non-linear evolution of the field before it decays \cite{Dimopoulos:2003ss,Enqvist:2005pg,Huang:2008zj,Enqvist:2009ww}. In this case one may find large negative values of $\fnl$ for specific model parameters and initial conditions.

Another distinctive aspect of the curvaton model is the possibility to leave residual isocurvature perturbations after the curvaton has decayed \cite{Mollerach:1989hu,Linde:1996gt,Lyth:2001nq}. Non-adiabatic perturbations on super-Hubble scales during inflation are necessary, but not sufficient, condition for the existence of primordial isocurvature perturbations after inflation. Whether or not residual isocurvature modes survive after the curvaton decays depends upon the reheating history \cite{Lyth:2003ip}. If all the curvaton decay products thermalise with vanishing chemical potential then the primordial perturbation must be adiabatic \cite{Weinberg:2003sw}. But if a baryon or lepton asymmetry has already been created, or the curvaton decay itself produces the asymmetry, or the dark matter has already decoupled, then a residual isocurvature perturbation, $S_X$, may be left, and it will be completely correlated (or anti-correlated) with the total density perturbation, $\zeta$.

For example, if the curvaton decay is the out-of-equilibrium process which breaks time-reversal invariance and violates the baryon number then the baryon number inherits the same density perturbation as the curvaton, $\zeta_B=\zeta_\chi$, to all orders~\cite{Langlois:2008vk}. In this case bounds on the amplitude of the linear primordial isocurvature perturbation \cite{Komatsu:2010fb} requires that the curvaton dominate the energy density when it decays, $r\simeq1$, so that $\zeta\simeq\zeta_B$ and, from Eq.~(\ref{defSB}), $S_B\ll\zeta_B$. In this case the intrinsic non-linearity parameter of the isocurvature perturbation is of order unity, and the contribution to the CMB bispectrum is suppressed relative to that from the total density perturbation.

A mixed curvaton-inflaton scenario yields a much richer phenomenology of adiabatic, non-adiabatic and non-Gaussianity of primordial perturbations \cite{Langlois:2004nn,Ferrer:2004nv,Lazarides:2004we,Ichikawa:2008iq,Langlois:2008vk,Lemoine:2009is}.

\subsection{Multiple-field inflation}

The curvaton is just one example of how fields other than a single inflaton driving inflation, could play a significant role in determining the primordial density perturbation after the end of inflation.
The presence of non-adiabatic perturbations on super-Hubble scales allows in principle for the evolution of the large-scale density perturbation after Hubble-exit \cite{Gordon:2000hv} and hence local non-Gaussianity in the primordial density perturbation some time after inflation.

One of the best studied models is double inflation with two massive but non-interacting fields \cite{Starobinsky:1986fxa,Langlois:1999dw}
where the non-linearity parameter though not necessarily slow-roll suppressed is not expected to be much larger than unity \cite{Rigopoulos:2005ae,Alabidi:2005qi,Vernizzi:2006ve}. Primordial isocurvature perturbations have also been studied in this model where one of the massive fields is identified as a dark matter candidate \cite{Polarski:1994rz,Langlois:1999dw,Langlois:2008vk}.

In general it seems surprisingly difficult to produce large non-linearity parameters during slow-roll inflation, but examples can be constructed~\cite{Byrnes:2008wi}. For more discussion see
the article by Tanaka {\em et al} in this volume~\cite{Tanaka:2010km}, and
the recent review by Byrnes and Choi~\cite{Byrnes:2010em}.
Non-Gaussianities could be produced by the breakdown of slow-roll due to features in the potential \cite{Chen:2006xjb} or particle production \cite{Langlois:2009jp}.

In inflationary models where more than one field remains light until the end of inflation it is necessary to consider possible evolution of the density perturbation on large scales at or after the end of inflation. The end of slow-roll inflation is an epoch at which the large-scale density perturbation may have a non-linear dependence upon non-adiabatic modes, especially it there is an abrupt change in the equation of state \cite{Bernardeau:2004zz,Lyth:2005qk,Salem:2005nd,Sasaki:2008uc}.

It is very natural to realise this possibility in hybrid inflation models \cite{Linde:1993cn} where inflation is ended by a tachyonic instability triggered in a waterfall field, leading to a rapid phase transition and decay of the false vacuum. In string theory models of inflation this may describe the collision of branes in a higher-dimensional space \cite{Lyth:2006nx}. The instability corresponds to a surface in field space, and if the inflaton trajectory is not orthogonal to this surface then non-adiabatic field perturbations, orthogonal to the trajectory~\cite{Gordon:2000hv}, are converted into density perturbations at the phase transition. In the simplest case, taking the extreme slow-roll limit and assuming instantaneous reheating, the energy density will be almost uniform on the spatial hypersurface corresponding to the phase transition, and thus the perturbed expansion to this surface, $\delta N$, can be identified with the primordial density perturbation, $\zeta$ in Eq.(\ref{defzeta}).

\subsection{Modulated decay}

A specific example where the primordial density perturbations may be produced just after the end of inflation is when the decay rate of the inflaton is a function of one or more moduli fields \cite{Dvali:2003em,Kofman:2003nx}. Because this occurs some time after Hubble exit during inflation, this produces local type non-Gaussianity \cite{Zaldarriaga:2003my,Vernizzi:2003vs,Kohri:2009ac}.

If we approximate the inflaton reheating by a sudden decay, as we did earlier for the curvaton decay, we find an analytic estimate of the non-linear density perturbation. In the case of modulated reheating, the decay occurs on a spatial hypersurface with variable local decay rate and hence local Hubble rate $H_d=\Gamma(\chi)$. Before inflaton decay, the oscillating field has a pressureless equation of state and there is no density perturbation perturbation. Setting $\zeta=0$ in Eq.~(\ref{defzeta}) thus gives the perturbed expansion on the decay hypersurface
\be
 \delta N_d = -\frac13 \ln \left( \frac{\rho_d}{\bar\rho_d} \right) \,.
\ee
After the decay the we have radiation with equation of state $p=\rho/3$ and hence from Eq.~(\ref{defzeta}) a density perturbation
\be
 \zeta = \delta N_d + \frac14  \ln \left( \frac{\rho_d}{\bar\rho_d} \right) \,.
\ee
Eliminating $\delta N_d$ and using the local Friedmann equation, $\rho\propto H^2$, to determine the local density in terms of the local decay rate, $H_d=\Gamma(\chi)$, we have
\be
 \zeta = -\frac16 \ln \left( \frac{\Gamma(\chi)}{\bar\Gamma} \right) \,.
 \ee
At first order we recover the linear relation \cite{Dvali:2003em,Vernizzi:2003vs}
\be
 \zeta_1 = -\frac16 \frac{\Gamma'}{\Gamma} \delta\chi \,.
\ee
At higher order we obtain \cite{Kohri:2009ac}
\bea
 \zeta_2 = \frac16 \left[ \left( \frac{\Gamma'}{\Gamma} \right)^2 - \frac{\Gamma''}{\Gamma} \right] \delta\chi^2 \,,\\
 \zeta_3 = \frac16 \left[ -2\left( \frac{\Gamma'}{\Gamma} \right)^3 + 3\frac{\Gamma''\Gamma'}{\Gamma^2} - \frac{\Gamma'''}{\Gamma} \right] \delta\chi^3
 \,,
 \eea
and hence
\bea
 \fnl = 5 \left( 1 - \frac{\Gamma''\Gamma}{\Gamma^{\prime2}} \right) \,,\\
 \gnl = \frac{50}{3} \left( 2 - 3\frac{\Gamma\Gamma''}{\Gamma^{\prime2}} + \frac{\Gamma'''}{\Gamma^{\prime3}} \right) \,,
\eea
and so on to higher order. Thus for an approximately linear modulation function $\Gamma(\chi)$ we have $\fnl\simeq5$ and $\gnl\simeq100/3$ while for $\Gamma\propto\chi^2$ we obtain $\fnl\simeq 5/2$ and $\gnl\simeq25/3$. Similar results are found if the mass of the inflaton decay products is modulated, so-called inhomogeneous mass domination \cite{Dvali:2003ar,Vernizzi:2003vs}.

Resonant decay or preheating may be sensitive to non-adiabatic modes \cite{Kolb:2004jm,Ackerman:2004kw,Kolb:2005ux,Byrnes:2005th,Matsuda:2006ee,Kohri:2009ac}. In these models observational constraints on primordial non-Gaussianity impose significant constraints on the allowed parameter values and initial conditions \cite{Byrnes:2008zz}. For a review of inflaton dynamics and reheating see Bassett et al \cite{Bassett:2005xm}.

\subsection{Ekpyrotic model}

There have been numerous attempts to construct alternatives to inflation as a model for the origin of primordial perturbations. Many of these exploit the similarities between an inflationary expansion and a collapse phase where the comoving Hubble scale decreases and thus quantum vacuum fluctuations evolve into the super-Hubble regime \cite{Gasperini:1993hu,Wands:1998yp}. As the Hubble rate grows during collapse, these models generally require some form of instability for the perturbations to grow on super-Hubble scales and maintain a scale-invariant spectrum.
For example, a simple collapse model driven by a scalar field with pressureless equation of state generates a scale-invariant spectrum for $\zeta$ on super-Hubble scales \cite{Wands:1998yp,Finelli:2001sr}. The non-Gaussianity in this has recently been calculated \cite{Cai:2009fn} and shown to yield $\fnl\sim1$, though not of the local form in Eq.~(\ref{originalPhi}).
Ultimately the four-dimensional low-energy effective theory must break down as the collapse rate approaches the Planck scale and this leaves some uncertainty about how the perturbations are transfered to the expanding hot big bang, but at least within the collapse phase it is possible to study the growth of perturbation $\zeta$ and its non-Gaussianity.

Recent attention has focussed on the ekpyrotic model \cite{Khoury:2001wf,Kallosh:2001ai}. In the 4D effective theory this corresponds to a cosmology driven by scalar fields with steep exponential potentials, $V_I(\varphi^I)\propto \exp(-c_I\varphi^I)$ (no sum) leading to an ultra-stiff equation of state $p/\rho\gg1$. The separate universe picture is an excellent approximation on super-Hubble scales during an ekpyrotic phase \cite{Lyth:2003im,Lehners:2009ja}. Thus, for fields with canonical kinetic Lagrangians, the local values of the fields perturbations at Hubble-exit have an approximately Gaussian distribution and set the initial conditions the subsequent local expansion history. As a result we expect the primordial non-Gaussianity to have a local form, and due to the strong self-interaction terms in steep potentials, we expect the non-linearities to become large~\cite{Creminelli:2007aq}.

The adiabatic mode during an ekpyrotic collapse leads to a steep blue spectrum for $\zeta$~\cite{Lyth:2001pf} but in the presence of two or more fields with steep exponential potentials it is possible to produce an almost scale-invariant spectrum of isocurvature perturbations~\cite{Notari:2002yc,Finelli:2002we,Lehners:2007ac,Buchbinder:2007ad,Creminelli:2007aq}. Consider the simple case of two fields with potential
\begin{equation}
 \label{simpleekpyrotic}
 V = -V_1 \exp \left(-c_1\varphi^{(1)}\right)  - V_2 \exp \left(-c_2\varphi^{(2)}\right) \,,
 \end{equation}
Performing a rotation in field-space this can be re-written as \cite{Koyama:2007mg}
\begin{equation}
 V=V_0\exp(-c\sigma) \left[ -1 -\frac{c^2}{2} s^2 - \frac{K_3c^3}{12\sqrt{2}}s^3 - \frac{K_4c^4}{96}s^4 \ldots \right] \,,
 \end{equation}
where the fast-roll parameter $c^{-2}=c_1^{-2}+c_2^{-2}$ and we identify $s$ as the tachyonic field direction which acquires an almost scale-invariant spectrum of perturbations about $s=0$ on super-Hubble scales, with spectral tilt
\be
 \label{ekpyrotictilt}
 \frac{d\ln {\cal P}}{d\ln k} = \frac{4}{c^2} \,.
 \ee
For an almost scale-invariant spectrum we require $c^2\gg1$.

Note that in the $\delta N$ formula the scale dependence of the primordial power spectrum (at leading order) follows directly from the scale-dependence of the initial field perturbations, $n_\zeta-1=4/c^2$, but the amplitude depends on the subsequent expansion history. In the ekpyrotic scenario different models have been proposed for the conversion of the isocurvature field perturbations to density perturbation $\zeta$. The tachyonic instability itself will lead to a phase transition to an ekpyrotic phase dominated by just one of the fields, $\varphi^{I}$~\cite{Koyama:2007ag}. In this case $N(\varphi^{(1)},\varphi^{(2)})$ can be calculated analytically for the simple model (\ref{simpleekpyrotic}) and we find \cite{Koyama:2007if,Lehners:2009ja}.
\be
\fnl \simeq -\frac{5}{12}c_I^2 \,, \qquad \gnl \simeq \frac{25}{108} c_I^4  \,.
\ee
Note that $c_I^2>c^2\gg1$ and therefore the non-linearity parameters are expected to be large. Tight observational bounds on a negative value for $\fnl>-10$ \cite{Komatsu:2010fb} require $c_I^2<24$ in this simple model and hence from Eq.~(\ref{ekpyrotictilt}) an unacceptably blue primordial spectral tilt $n_\zeta-1>0.16$. An acceptable tilt and non-Gaussianity would require modifications to the simple potential Eq.~(\ref{ekpyrotictilt}) which itself could trigger a transition and a isocurvature-curvature conversion. The amplitude and sign of the non-linearity parameters then becomes model-dependent but one generally finds $\fnl^2\sim|\gnl|\sim c^2$ \cite{Buchbinder:2007at}.

Alternative conversion mechanisms would produce a different amplitude of primordial perturbations and hence non-Gaussianity. In models where the conversion happens at a reflection in field space during a kinetic dominated phase give estimates for the non-linearity parameters \cite{Lehners:2009ja}.
\be
\fnl \sim - \frac{3}{2\sqrt{2}} K_3 c + 5 \,, \qquad \gnl \simeq -20 c^2 \left[ 1 - \frac{K_3^2}{32} - \frac{K_4}{24} \right] \,.
\ee
$K_3$ and $K_4$ are expected to be of order unity. Again the non-linearity parameters are expected to be large with $|\gnl|\sim\fnl^2\sim c^2$. In this case we see that even if $K_3$ is small, such that $|\fnl|\sim1$ [corresponding to $c_1^2-c_2^2\ll 1/c$ in Eq.~(\ref{simpleekpyrotic})], then we would expect $|\gnl|$ to be large, unless $K_4$ is also unexpectedly small.
We note the further possibility that isocurvature field fluctuations are converted to density perturbations only after the bounce \cite{Battefeld:2007st}, as in curvaton or modulated decay models.

\section{Conclusions}

The distribution of primordial density perturbations contains more information than just the power-spectrum. Among a plethora of different possible forms of non-Gaussianity the local models provide a clearly defined class which arise naturally from the evolution of the density perturbation on super-Hubble scales from initially Gaussian vacuum fluctuations during inflation. In particular the $(2+n)$-point function of a non-Gaussian distribution generated by a local function of a single Gaussian field can be described at leading order by the two-point function and $n$ non-linearity parameters, $\fnl$, $\gnl$, etc.

Non-Gaussianities have the potential to reveal the physical interactions at work in the very early Universe.
For adiabatic perturbations during inflation the non-linearity parameters are related to the scale-dependence of the power spectrum \cite{Maldacena:2002vr,Creminelli:2004yq} and hence must be small. Thus any detection of local non-Gaussianity of the primordial density perturbation would be evidence of non-adiabatic perturbations on super-Hubble scales, e.g., the presence of multiple light fields during slow-roll inflation.

Current observational limits on $\fnl$ in the single-field local model come from combining WMAP cosmic microwave background bounds ($-10<\fnl<74$ \cite{Komatsu:2010fb}) with large-scale structure ($-29<\fnl<70$ \cite{Slosar:2008hx}), yielding $-5<\fnl<70$ at 95\% confidence limit~\cite{Komatsu:2010fb}. Recent work also places a bound $-3.80<\gnl/10^6<3.88$ from the WMAP 5-year data \cite{Smidt:2010sv}. Care should be taken when applying these bounds to non-local or even multi-variate local models. CMB bounds are obtained using matched filtering techniques to construct an optimal estimator for $\fnl$ defined by Eq.~(\ref{originalPhi}). It estimates the amplitude of a second-order primordial perturbation which has a specified correlation with the first-order perturbation. On the other hand bounds from large scale structure arise from the form of the (general relativistic) Poisson equation \cite{Wands:2009ex} which implies that a local form for the primordial Newtonian potential requires a non-local form for the comoving density contrast, leading to a distinctive scale-dependent bias \cite{Dalal:2007cu}.

There is great potential for future discovery coming from future CMB missions such as ESA's Planck satellite which has the potential to bound $|\fnl|<5$ and large-scale structure surveys, including 21cm line radio surveys \cite{Komatsu:2009kd}. There is also plenty of scope for further theoretical developments given the data that already exists to test non-Gaussian models. New theoretical templates will need to be derived for optimised constraints to be placed on alternative theoretical models of non-Gaussianity. At some point primordial non-Gaussianity will be discovered as it inevitably arises from the non-linear evolution of density perturbations and the last-scattering of the CMB photons \cite{Bartolo:2004if,Pitrou:2010sn}, and it will then be a question of using all the available data to disentangle the different contributions.

\section*{Acknowledgements}

I am grateful to my many collaborators for the work on which this highly subjective review is based, and especially Chris Byrnes, David Langlois, David Lyth, Karim Malik, Misao Sasaki, Jussi Valiviita and Filippo Vernizzi. DW is supported by the STFC and is grateful to the Yukawa Institute of Theoretical Physics, Kyoto, for their hospitality.

\appendix
\section{Sign conventions}

We collect together here definitions (at linear order) and sign conventions for the gauge-invariant curvature perturbations used in the literature.

Given the FRW line element with scalar metric perturbations 
\bea
ds^2 = a^2(\eta) \left[ - (1+2\A)dt^2 + 2\partial_i \B dx^i d\eta +
 \right. \nonumber\\
\qquad \qquad \qquad \left. \left\{ (1+2\C)\delta_{ij} + 2 \partial_i\partial_j \E \right\} dx^i dx^j \right] \,,
\eea
the intrinsic spatial curvature is given by the Ricci scalar on constant-$\eta$ hypersurfaces
\be
^{(3)}R = - \frac{4}{a^2} \partial^2 \C \,.
\ee

We can define three gauge invariant curvature perturbations at first-order:
\begin{enumerate}
\item
uniform-density curvature perturbation
\be
 \zeta \equiv \C + \frac{\delta\rho}{3(\rho+P)} \,,
\ee
\item
comoving-orthogonal curvature perturbation
\be
 \R \equiv \C - \frac{\H\delta\varphi}{\varphi'} \,,
\ee
\item
longitudinal-gauge curvature perturbation
\be
 \Phi \equiv \C + \H (\B-\E') \,,
\ee
\end{enumerate}
where primes denote derivatives with respect to the conformal time $\eta$ and $\H=a'/a$.
Note that in the longitudinal gauge the curvature perturbation can be simply related to the Newtonian potential.

The sign of the metric perturbation at the gauge-invariant variables formed from it are arbitrary and different authors have adopted different signs and notations. A different choice of sign for $\Phi$ in Eq.(\ref{bispectrum}) then leads to a different choice of sign for $\fnl$ in Eq.(\ref{deffnl}). This too is purely conventional, but once this choice is specified observations are sensitive to the sign of the resulting $\Phi$ and hence the sign of $\fnl$. Using the sign conventions in this paper, a CMB sky with positive $\fnl$ has more cold spots, and one with negative $\fnl$ has more hot spots.

Different notations and sign conventions used in specific papers are listed in table~A1.

\begin{table}

\begin{tabular}{|c|c|c|c|c|}
\hline
                & $\zeta$   & $\R$  & $\Phi$    & $\fnl$ \\
\hline
\hline
this article    & + & $+$ & $+$ & +\\
\hline
Malik \& Wands \cite{Malik:2008im} & + & $-$ & $-\Psi$  & + \\
\hline
Bassett, Tsujikawa \& Wands \cite{Bassett:2005xm} & + & $-$ & $-\Psi$ & $-$ \\
\hline
Lyth \& Rodriguez \cite{Lyth:2005fi} & $+$ & . & $-$ & $-$ \\
\hline
Lyth, Malik \& Sasaki (sections 3 \& 4 only)
\cite{Lyth:2004gb} & $-$ & . & . & . \\
\hline
Maldacena \cite{Maldacena:2002vr} & . & $+\zeta$ & $-$ & $-$ \\
\hline
Komatsu \& Spergel \cite{Komatsu:2001rj} & . & . & + & + \\
\hline
Liddle \& Lyth \cite{Liddle:2000cg} & . & $+$ & $-$ & . \\
\hline
Mukhanov, Feldman \& Brandenberger \cite{Mukhanov:1990me} & . & $-\zeta$ & $-\Psi$ & . \\
\hline
Kodama \& Sasaki \cite{Kodama:1985bj} & . & + & + & . \\
\hline
Bardeen \cite{Bardeen:1980kt} & . & $+\phi_m$ & $+\Phi_H$ & . \\
\hline
 \end{tabular}

\caption{Summary of sign conventions and alternative notations used in specific papers.}

\end{table}

\end{document}